\documentclass{svjour3}
\smartqed
\usepackage{graphicx}
\usepackage{bm}
\usepackage{amssymb}
\usepackage{amsmath}

\begin{document}

\title{Innovative methods of correlation and orbit determination for
  space debris}

\author{Davide Farnocchia \and Giacomo Tommei \and Andrea Milani \and
  Alessandro Rossi}

\institute{D. Farnocchia, G. Tommei, A. Milani \at Department of
  Mathematics, University of Pisa, Largo
  Bruno Pontecorvo 5, 56127 Pisa, Italy\\
  \email{farnocchia@mail.dm.unipi.it} \and G. Tommei \at
  \email{tommei@dm.unipi.it} \and A. Milani \at
  \email{milani@dm.unipi.it} \and A. Rossi \at
  ISTI/CNR, Via Moruzzi 1, 56124 Pisa, Italy\\
  \email{alessandro.rossi@isti.cnr.it} }

\date{Received: date / Accepted: date}

\maketitle

\begin{abstract}
  We propose two algorithms to provide a full preliminary orbit of an
  Earth-orbiting object with a number of observations lower
  than the classical methods, such as those by Laplace and
  Gauss. The first one is the Virtual debris algorithm, based upon the
  admissible region, that is the set of the unknown quantities
  corresponding to possible orbits for objects in Earth
  orbit (as opposed to both interplanetary orbits and ballistic
  ones). A similar method has already been successfully used in recent
  years for the asteroidal case. The second algorithm uses the
  integrals of the geocentric 2-body motion, which must have the
  same values at the times of the different observations for a common
  orbit to exist. We also discuss how to account for the perturbations
  of the 2-body motion, e.g., the $J_2$ effect.

\end{abstract}

\section{Introduction}

The near-Earth space, filled by more than 300000 artificial debris
particles with diameter larger than 1 cm, can be divided into three
main regions: the Low Earth Orbit (LEO), below about 2000 km, the
Medium Earth Orbit (MEO), above 2000 km and below 36000 km, and the
Geosynchronous Earth Orbit (GEO) at about 36000 km of altitude.
Currently the orbits of more than 12000 objects larger than about 10
cm are listed in the so called Two Line Elements (TLE) catalogue. To
produce and maintain such a catalogue a large number of optical and
radar observations are routinely performed by the United States Space
Surveillance Network. Nowadays also Europe has launched its Space
Situational Awareness (SSA) initiative aimed to increase the knowledge
of the circumterrestrial environment. In this context the availability
of efficient methods and algorithms for accurate orbit determination
is extremely important.

Given two or more sets of observations, the main problem is how to
identify which separate sets of data belong to the same physical
object (the so-called \emph{correlation} problem). Thus the orbit
determination problem needs to be solved in two stages: first
different sets of observations need to be correlated, then an orbit
can be determined; this combined procedure is called \emph{linkage} in
the literature \cite{milani_recovery}.

In this paper we describe two different linkage methods, for both
optical and radar data. By using the attributable vector
(Sec.~\ref{sec:att}) we summarize the information contained in either
optical or radar data. In Sec.~\ref{sec:ad_reg} we describe the
admissible region and the Virtual debris algorithm \cite{tommei} and
we propose a general scheme to classify observed
objects. Sec.~\ref{sec:kepint} deals with the Keplerian integrals
method, first introduced by \cite{gronchi} for the problem of asteroid
orbit determination. Furthermore, the inclusion of the effect due to
the non-spherical shape of the Earth is discussed. Finally, in
Sec.~\ref{sec:correl_confirm}, a sketch of the general procedure for
the full process of correlation of different observations is outlined.

\section{Observations and attributables}
\label{sec:att}

Objects in LEO are mostly observed by radar while for MEOs and GEOs
optical sensors are used. In both cases, the batches of observations
which can be immediately assigned to a single object give us a set of
data that can be summarized in an attributable, that is a
4-dimensional vector. To compute a full orbit, formed by 6 parameters,
we need to know 2 further quantities.

Thus the question is the identification problem, also called
correlation in the debris context: given 2 attributables at different
times, can they belong to the same orbiting object? And if this is the
case, can we find an orbit fitting both data sets?


Let $(\rho,\alpha,\delta) \in \mathbb R^+ \times [0,2\pi) \times
(-\pi/2, \pi/2)$ be spherical coordinates for the topocentric position
of an Earth satellite. The angular coordinates $(\alpha,\delta)$ are
defined by a topocentric reference system that can be arbitrarily
selected.  Usually, in the applications, $\alpha$ is the right
ascension and $\delta$ the declination with respect to an equatorial
reference system (e.g., J2000). The values of range $\rho$ and range
rate $\dot\rho$ are not measured.

We shall call \textbf{optical attributable} a vector
\[ 
{\cal A}_{opt}=(\alpha,\delta,\dot\alpha,\dot\delta) \in [0,2\pi)
\times (\pi/2,\pi/2) \times \mathbb R^2\, ,
\]
representing the angular position and velocity of the body at a time
$t$ in the selected reference frame.


Active artificial satellites and space debris can also be observed by
radar: because of the $1/\rho^4$ dependence of the signal to noise for
radar observations, range and range-rate are currently measured only
for debris in LEO. When a return signal is acquired, the antenna
pointing angles are also available. Given the capability of modern
radars to scan very rapidly the entire visible sky, radar can be used
to discover all the debris above a minimum size while they are visible
from an antenna, or a system of antennas.

When a radar observation is performed we assume that the measured
quantities (all with their own uncertainty) are the range, the range
rate, and also the antenna pointing direction, that is the debris
apparent position on the celestial sphere, expressed by two angular
coordinates such as right ascension $\alpha$ and declination
$\delta$. The time derivatives of these angular coordinates,
$\dot\alpha$ and $\dot\delta$, are not measured.

We define \textbf{radar attributable} a vector
\[ 
{\cal A}_{rad}=(\alpha,\delta,\rho,\dot\rho) \in [-\pi,\pi) \times
(-\pi/2,\pi/2) \times \mathbb R^+ \times \mathbb R\, ,
\]
containing the information from a radar observation, at the receive time $t$.


To define an orbit given the attributable ${\cal A}$ we need to find
the values of two unknowns quantities (e.g., $\rho$ and $\dot\rho$ in
the optical case, $\dot\alpha$ and $\dot\delta$ in the radar case),
that, together with the attributable, give us a set of \textbf{attributable
orbital elements}:
\[
X = [\alpha,\delta,\dot\alpha,\dot\delta,\rho,\dot\rho]
\]
at a time $\bar t$, computed from $t$ taking into account the
light-time correction: $\bar t=t-\rho/c$. The Cartesian geocentric
position and velocity $(\mathbf r,\dot{\mathbf r})$ can be obtained,
given the observer geocentric position $\mathbf q$ at time $t$, by using the
unit vector $\hat{\bm\rho}=
(\cos\alpha\cos\delta,\sin\alpha\cos\delta,\sin\delta)$ in the
direction of the observation:
\[
\mathbf r=\mathbf q+\rho\hat{\bm\rho}\ \ ,\ \ \dot{\mathbf
  r}=\dot{\mathbf
  q}+\dot\rho\hat{\bm\rho}+\rho\frac{d\hat{\bm\rho}}{dt}\ \ \
, \ \ \
\frac{d\hat{\bm\rho}}{dt}=\dot\alpha\hat{\bm\rho}_\alpha+
\dot\delta\hat{\bm\rho}_\delta\ ,
\]
\[
\hat{\bm\rho}_\alpha=(-\sin\alpha\cos\delta,\cos\alpha\cos\delta,0)\,
,\ \ \
\hat{\bm\rho}_\delta=(-\cos\alpha\sin\delta,-\sin\alpha\sin\delta,
\cos\delta)\, .
\]

\section{Admissible region theory}
\label{sec:ad_reg}

Starting from an attributable, we would like to extract sufficient
information from it in order to compute preliminary orbits: we shall
use the admissible region tool, as described in \cite{tommei}. For
easy of reading we recall here the basic steps of the theory.

The admissible region replaces the conventional confidence region as
defined in the classical orbit determination procedure. The main
requirement is that the geocentric energy of the object is negative,
that is the object is a satellite of the Earth.

\subsection{Optical admissible region}

Given the geocentric position $\mathbf r$ of the debris, the
geocentric position $\mathbf q$ of the observer, and the topocentric
position $\bm\rho$ of the debris we have $\mathbf r=\bm\rho+\mathbf
q$. The energy (per unit of mass) is given by
\[ 
{\cal E}(\rho,\dot\rho)=\frac{1}{2}||\dot{\mathbf
  r}(\rho,\dot\rho)||^2-\frac{\mu}{||\mathbf r(\rho)||}\, ,
\]
where $\mu$ is the Earth gravitational parameter. Then a definition of
admissible region such that only satellites of the Earth are allowed
includes the condition
\begin{equation}\label{negative_energy}
{\cal E}(\rho,\dot\rho)\leq 0
\end{equation}
that could be rewritten as
\[
2{\cal E}(\rho,\dot\rho)=\dot\rho^2+c_1\dot\rho+T(\rho)-\frac{2\mu}
{\sqrt{S(\rho)}}\leq 0\, ,
\]
\[
T(\rho)=c_2\rho^2+c_3\rho+c_4\ \ ,\ \ S(\rho)=\rho^2+c_5\rho+c_0
\]
and coefficients $c_i$ depending on the attributable \cite{tommei}:
\begin{align*}
  c_0&=||\mathbf q||^2\, ,&\ \ c_1&=2\,\dot{\mathbf q}
  \cdot\hat{\bm\rho}\, ,&\ \ c_2&=\dot\alpha^2\cos^2\delta+
  \dot\delta^2=\eta^2\, ,\\
  c_3&=2\,(\dot\alpha\,\dot{\mathbf q}\cdot\hat{\bm\rho}_\alpha+
  \dot\delta\,\dot{\mathbf q}\cdot\hat{\bm\rho}_\delta)\, ,&\ \
  c_4&=||\dot{\mathbf q}||^2\, ,&\ \ c_5&=2\,\mathbf q
  \cdot\hat{\bm\rho}\, ,
\end{align*}
where $\eta$ is the \emph{proper motion}.  In order to obtain real
solutions for $\dot\rho$ the discriminant of $2{\cal E}$ (polynomial
of degree 2 in $\dot\rho$) must be non-negative:
\[
\Delta=\frac{c_1^2}{4}-T(\rho)+\frac{2\mu}{\sqrt{S(\rho)}}\geq 0\, .
\]
This observation results in the following condition on $\rho$:
\begin{equation}\label{disc_condition}
  \frac{2\mu}{\sqrt{S(\rho)}}\geq Q(\rho)=c_2\rho^2+c_3\rho+\gamma\ \ 
  , \ \ \gamma=c_4-\frac{c_1^2}{4}\ .
\end{equation}
Condition (\ref{disc_condition}) can be seen as an inequality
involving a polynomial $V(\rho)$ of degree 6:
\[
V(\rho)=Q^2(\rho)S(\rho)\leq 4\mu^2\, .
\]
Studying the polynomial $V(\rho)$ and its roots, as done by
\cite{milani_adreg}, the conclusion is that the region of
$(\rho,\dot\rho)$ such that condition (\ref{negative_energy}) is
satisfied can admit more than one connected component, but it has at
most two. In any case, in a large number of numerical experiments with
objects in Earth orbit, we have not found examples with two connected
components.

\begin{figure*}
  \includegraphics[width=0.75\textwidth]{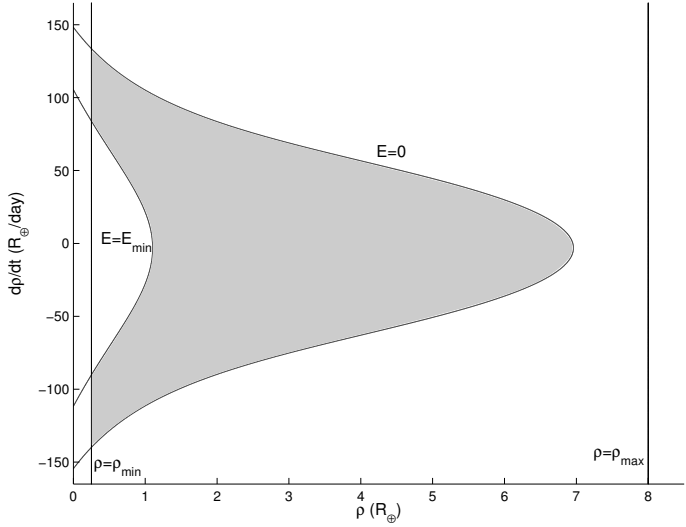}
  \caption{An example of admissible region, optical case, in the
    $(\rho,\dot\rho)$ plane. The region (painted in grey) is bounded
    by two level curves of the energy, ($E=E_{min}$) and ($E=0$), and
    by the two conditions on the topocentric distance
    ($\rho=\rho_{min}$ and $\rho=\rho_{max}$).}
\label{fig:adregion_optical1}
\end{figure*}

The admissible region needs to be compact in order to have the
possibility to sample it with a finite number of points, thus a
condition defining an inner boundary needs to be added. The choice for
the inner boundary depends upon the specific orbit determination task:
a simple method is to add constraints
$\rho_{min}\leq\rho\leq\rho_{max}$ allowing, e.g., to focus the search
of identifications to one of the three classes LEO, MEO and
GEO. Another natural choice for the inner boundary is to take
$\rho\geq h_{atm}$ where $h_{atm}$ is the thickness of a portion of
the Earth atmosphere in which a satellite cannot remain in orbit for a
significant time span. As an alternative, it is possible to constrain
the semimajor axis to be larger than $R_\oplus+h_{atm}=r_{min}$, and
this leads to the inequality
\begin{equation}\label{E_min}
{\cal E}(\rho,\dot\rho)\geq -\frac{\mu}{2r_{min}}={\cal E}_{min}\ ,
\end{equation}
which defines another degree six inequality with the same coefficients
but for a different constant term. The qualitative structure of the
admissible region is shown in Fig.~\ref{fig:adregion_optical1}.

Another possible way to find an inner boundary is to exclude
trajectories impacting the Earth in less than one revolution, that is
to use an inequality on the perigee $r_P$, already proposed in
\cite{maruskin}:
\begin{equation}\label{pericenter1}
r_P=a(1-e)\geq r_{min}.
\end{equation}
Note that this condition naturally implies (\ref{E_min}) and $\rho\geq
h_{atm}$.
To analytically develop the inequality (\ref{pericenter1}) we use the
2-body formulae involving the angular momentum:
\begin{equation}\label{angmom_opt}
\mathbf c(\rho,\dot\rho)=\mathbf r\times\dot{\mathbf r}=\mathbf
D\dot\rho+\mathbf E\rho^2+\mathbf F\rho+\mathbf G
\end{equation}
\begin{align*}
  \mathbf D&=\mathbf q \times\hat{\bm\rho},&\ \ \mathbf
  E&=\hat{\bm\rho}\times(\dot\alpha\hat{\bm\rho}_\alpha+
  \dot\delta\hat{\bm\rho}_\delta),\\
  \mathbf F&=\hat{\mathbf q}\times(\dot\alpha\hat{\bm\rho}_\alpha+
  \dot\delta\hat{\bm\rho}_\delta)+\dot{\mathbf q}\times\hat{\bm\rho},
  &\ \ \mathbf G&=\mathbf q \times\dot{\mathbf q}
\end{align*}
and substituting in (\ref{pericenter1}) we obtain:
\begin{equation}\label{pericenter2}
  \sqrt{1+\frac{2{\cal E}||\mathbf c||^2}{\mu^2}}\leq 1+\frac{2{\cal
      E} r_{min}}{\mu}\, .
\end{equation}
Since the left hand side is $e\geq 0$, we need to impose
$1+{2{\cal E}r_{min}}/{\mu}\geq 0$: this is
again $a\geq r_{min}$. By squaring (\ref{pericenter2}) we obtain:
\[
||\mathbf c||^2\geq 2r_{min} (\mu+{\cal E}r_{min})\, .
\]
The above condition is an algebraic inequality in the variables
$(\rho,\dot\rho)$:
\begin{equation}\label{pericenter3}
  (r_{min}^2-||\mathbf D||^2)\dot\rho^2-P(\rho)\dot\rho-U(\rho)+
  r_{min}^2T(\rho)-\frac{2r_{min}^2\mu}{\sqrt{S(\rho)}}\leq 0\ ,
\end{equation}
\begin{eqnarray*}
  P(\rho)&=&2\mathbf D\cdot\mathbf E\rho^2+2\mathbf 
  D\cdot\mathbf F\rho+2\mathbf D\cdot\mathbf G -r_{min}^2c_1\, ,\\
  U(\rho)&=&||\mathbf E||^2\rho^4+2\mathbf E\cdot\mathbf F\rho^3+
  (2\mathbf E\cdot\mathbf G+||\mathbf F||^2)\rho^2+
  2\mathbf F\cdot\mathbf G\rho+||\mathbf G||^2-2r_{min}\mu\, .
\end{eqnarray*}
The coefficient of $\dot\rho^2$ is positive, thus to obtain real
solutions for $\dot\rho$ the discriminant of (\ref{pericenter3}) must
be non negative:
\[
\Delta_P =P^2(\rho)+4(r_{min}^2-||\mathbf D||^2)\left(U(\rho)+
r_{min}^2T(\rho)+\frac{2r_{min}^2\mu}{\sqrt{S(\rho)}}\right)\geq 0\, .
\]
This condition is equivalent to the following:
\begin{equation}\label{ineq_peri}
\frac{2\mu}{\sqrt{S(\rho)}}\geq W(\rho)=-\frac{4(r_{min}^2-
  ||\mathbf D||^2)(U(\rho)+r_{min}^2T(\rho))
  +P^2(\rho)}{4r_{min}^2(r_{min}^2-||\mathbf D||^2)}\, .
\end{equation}
Note that the inequality (\ref{ineq_peri}) is similar to
(\ref{disc_condition}). However, in this case, the function in the
right hand side is much more complicated, and there is no easy way
to use the condition (\ref{pericenter1}) to explicitly describe the
boundary of the admissible region; e.g., we do not have a rigorous
bound on the number of connected components.
 This condition (\ref{pericenter1}) will be used only
{\it a posteriori} as a filter (Sec.~\ref{sec:vda}).

\begin{figure*}[h]
  \includegraphics[width=0.75\textwidth]{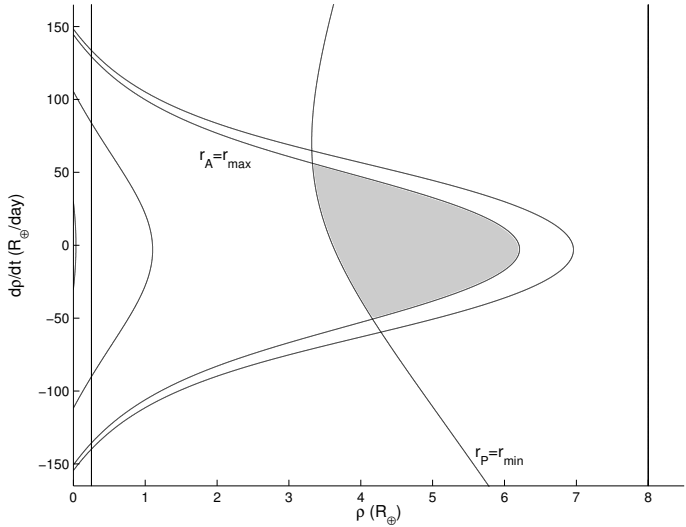}
  \caption{The same example of Fig.~\ref{fig:adregion_optical1},
    with the two further conditions on the pericenter ($r_P\geq
    r_{min}$) and the apocenter ($r_A\leq r_{max}$) distances.}
  \label{fig:adregion_optical2}
\end{figure*}

Fig.~\ref{fig:adregion_optical2} shows also this inner boundary;
note that the boundaries of the regions defined by (\ref{E_min}) and
by $\rho\geq h_{atm}$ are also plotted in the figure, but these
constraints are not necessary.
We have also plotted an alternative outer boundary constraining the
apocenter $r_A$ at some large value $r_{max}$:
\[
\begin{cases}
\displaystyle{\cal E}\leq-\frac{\mu}{2r_{max}}\\
\stackrel{}{||\mathbf c||^2\geq 2r_{max}(\mu+{\cal E}r_{max})}
\end{cases}
\ ;
\]
this outer boundary can be used in the same way, as an {\it a
  posteriori} filter.

\subsection{Radar admissible region}

Given a radar attributable ${\cal A}_{rad}$, we define as radar
admissible region for a space debris the set of values of
$(\dot\alpha,\dot\delta)$ such that
\begin{equation}\label{energy_rad} 
{\cal E}(\alpha,\delta)=z_1\dot\alpha^2+z_2\dot\delta^2+
z_3\dot\alpha+z_4\dot\delta+z_5\leq 0\, ,
\end{equation}
where $z_{ij}$ depend on the attributable \cite{tommei}:
\begin{align*}
z_1&=\rho^2\cos^2\delta\, ,&\ \ z_2&=\rho^2\, ,&\ \ 
z_3&=\rho\,\dot{\mathbf q}\cdot\hat{\bm\rho}_\alpha/2\, ,\\
z_4&=\rho\,\dot{\mathbf q}\cdot\hat{\bm\rho}_\delta/2\, ,&\ \
z_5&=\dot\rho^2+c_1\dot\rho+c_4-\frac{2\mu}{\sqrt{S(\rho)}}\, .&
\end{align*}
The boundary of the admissible region is then given by ${\cal
  E}(\dot\alpha,\dot\delta)=0$ and this equation represents an ellipse
with its axes aligned with the coordinate axes in the
$(\dot\alpha,\dot\delta)$ plane. Actually, in a plane
$(\dot\alpha\cos\delta,\dot\delta)$, with the axes rescaled according
to the metric of the tangent plane to the celestial sphere, the curves
${\cal E}(\dot\alpha,\dot\delta)=constant$ are circles.

\begin{figure*}
  \includegraphics[width=0.75\textwidth]{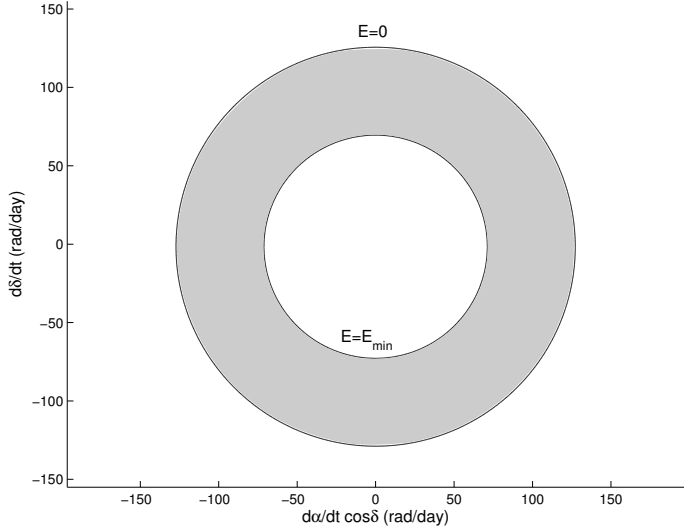}
  \caption{An example of admissible region, radar case, in the
    $(\dot\alpha\cos\delta,\dot\delta)$ plane. The region (painted in
    grey) is the circular annulus bounded by the two level curves of
    the energy ($E=E_{min}$) and ($E=0$).}
  \label{fig:adregion_radar1}
\end{figure*}

The region defined by negative geocentric energy, the inside of a
circle, is a compact set, and the problem of defining an inner
boundary is much less important than in the optical attributable
case. Anyway, it is possible to define an inner boundary by
constraining the semimajor axis $a>r_{min}$, that is by
eq. (\ref{E_min}), resulting in a concentric inner circle, thus in an
admissible region forming a circular annulus (see
Fig.~\ref{fig:adregion_radar1}).

It is also possible to exclude the ballistic trajectories by imposing
the condition (\ref{pericenter1}) in which $\dot\alpha,\dot\delta$ are
to be considered as variables. The angular momentum is given by
\begin{equation}\label{angmom_rad}
\mathbf c=\mathbf r\times\dot{\mathbf r}=\mathbf A\dot\alpha+\mathbf
B\dot\delta+\mathbf C\ ,
\end{equation}
\[
\mathbf A=\rho\,\mathbf r\times\hat{\bm\rho}_\alpha\, ,\ \ 
\mathbf B=\rho\,\mathbf r\times\hat{\bm\rho}_\delta\, ,\ \ 
\mathbf C=\mathbf r\times\dot{\mathbf q}+
\dot\rho\,\mathbf q\times\hat{\bm\rho}\, .
\]
The condition on the pericenter is expressed by a polynomial
inequality of degree 2:
\[
w_1\dot\alpha^2+w_2\dot\alpha\dot\delta+w_3\dot\delta^2
+w_4\dot\alpha+w_5\dot\delta+w_6\geq 0\ ,
\]
\begin{align*}
  w_1&=||\mathbf A||^2-2r_{min}^2z_1,&
  w_3&=||\mathbf B||^2-2r_{min}^2,&
  w_5&=2(\mathbf B\cdot\mathbf C-r_{min}^2z_4),\\
  w_2&=2\mathbf A\cdot\mathbf B,&
  w_4&=2(\mathbf A\cdot\mathbf C-r_{min}^2z_3),&
  w_6&=||\mathbf C||^2-2r_{min}(r_{min}z_5+\mu).
\end{align*}

\begin{figure*}
  \includegraphics[width=0.75\textwidth]{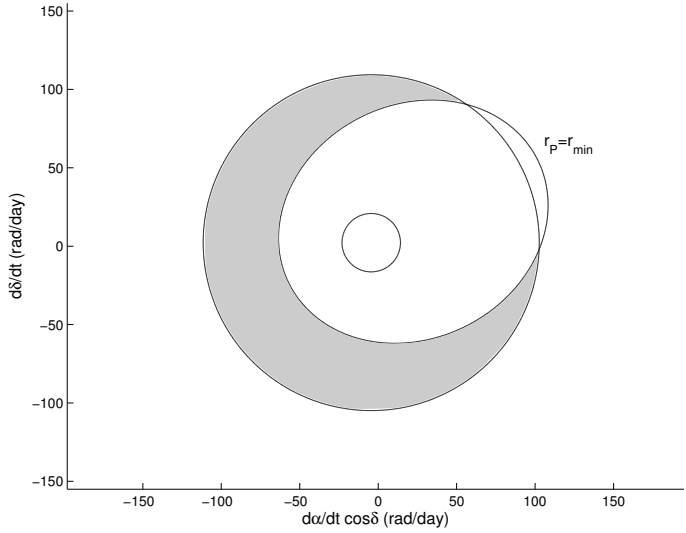}
  \caption{An example of admissible region, with the further condition
    on the pericenter distance ($r_P\geq r_{min}$), bounded by an
    ellipse.}
  \label{fig:aregion_radar_2}
\end{figure*}

\begin{figure*}
  \includegraphics[width=0.75\textwidth]{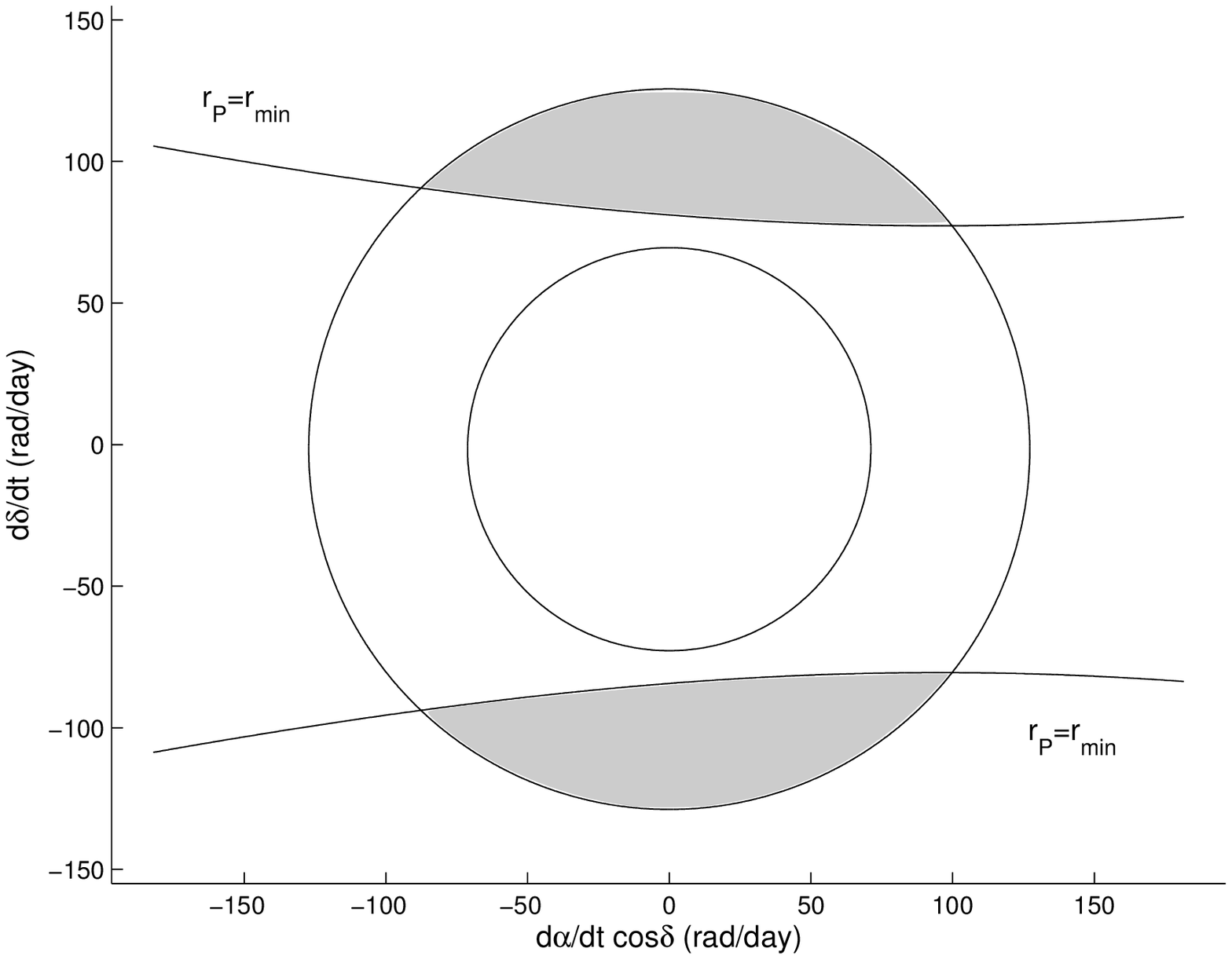}
  \caption{An example of admissible region, with the further condition
    on the pericenter distance ($r_P\geq r_{min}$), bounded by an
    hyperbola.}
\label{fig:aregion_radar_3}
\end{figure*}

Thus the admissible region can be geometrically described as a region
bounded by three conics: the first two are concentric circles, the
third one can be either an ellipse or an hyperbola (depending on the
sign of $w_1w_3-w_2^2/4$), with a different center and different
symmetry axes. Fig.~\ref{fig:aregion_radar_2} and
\ref{fig:aregion_radar_3} show the possible qualitatively different
cases.

\begin{figure*}[p]
  \includegraphics[width=0.75\textwidth]{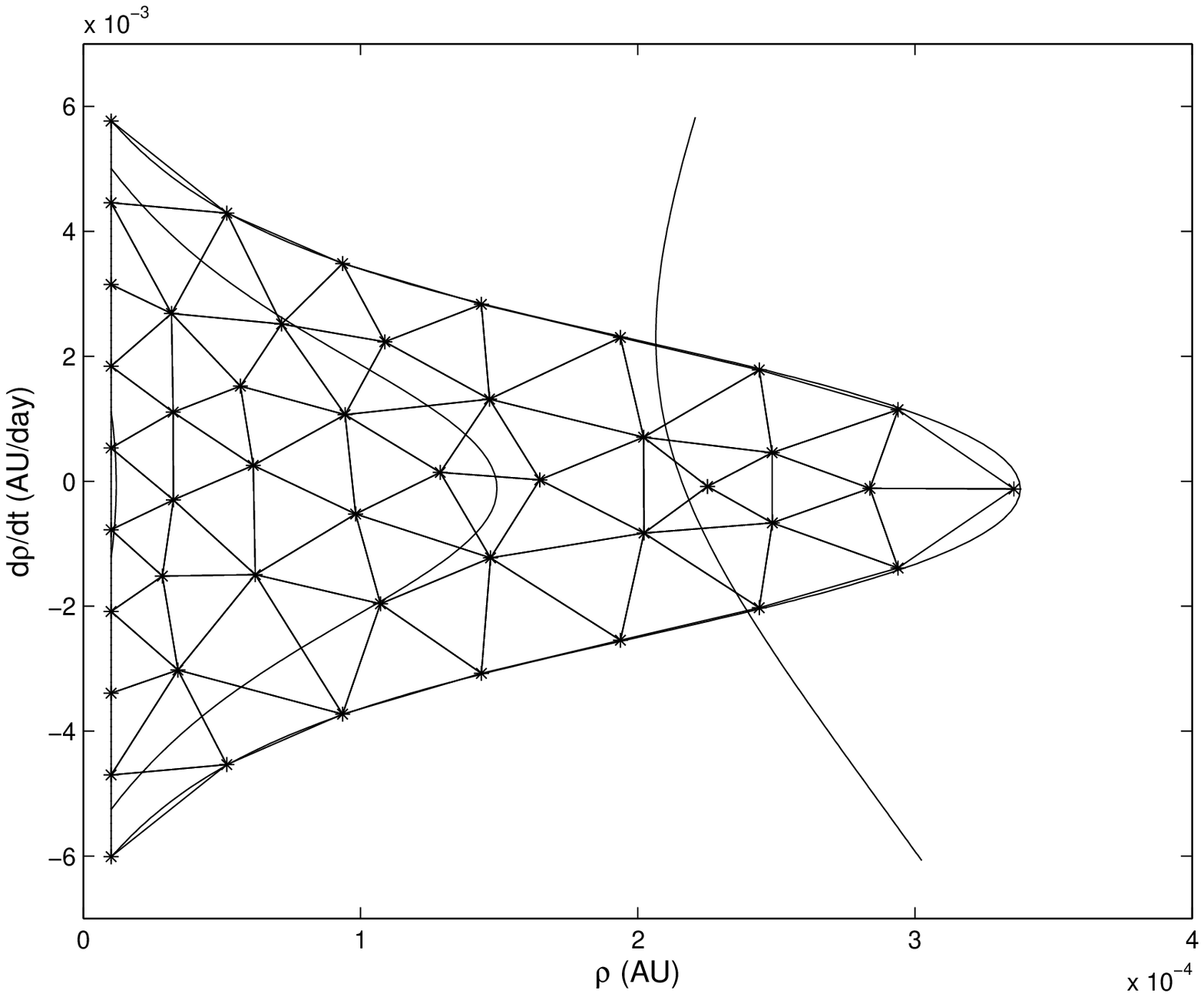}
  \caption{An example of admissible region, defined by imposing
    negative geocentric energy, for an optical attributable, with the
    Delaunay triangulation. The nodes of the triangulation
    corresponding to the ballistic trajectories (on the left of the
    curve cutting the outer part of the triangulation) can be
    discarded.}
  \label{fig:triang}

  \includegraphics[width=0.75\textwidth]{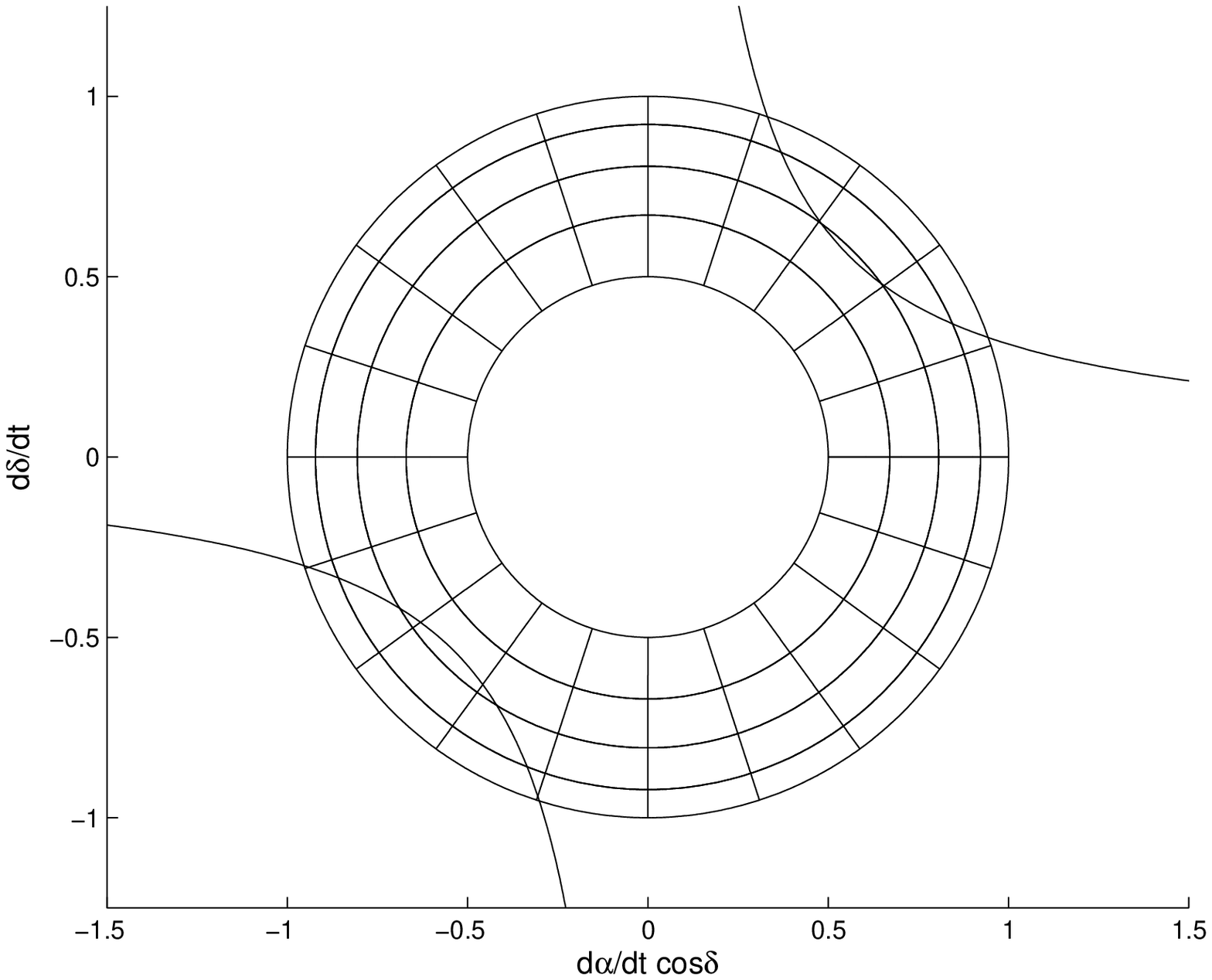}
  \caption{An example of admissible region, defined by imposing ${\cal
      E}_{min}\leq {\cal E}\leq 0$, for a radar attributable, with the
    cobweb sampling. The nodes of the cobweb corresponding to the
    ballistic trajectories (between the two branches of the hyperbola)
    can be discarded.}
  \label{fig:cobweb}
\end{figure*}

\subsection{Virtual debris algorithm}
\label{sec:vda}

The admissible region can be used to generate a swarm of virtual
debris: we sample it using the Delaunay triangulation
\cite{milani_adreg} for the optical case and the cobweb \cite{tommei}
for the radar case, as shown in Fig.~\ref{fig:triang} and \ref{fig:cobweb}. The condition
on the pericenter is not used at this step, because we could lose some
important geometrical properties: this condition is used as
filter, the nodes with a low pericenter are discarded.

The idea is to generate a swarm of virtual debris $X_i$, corresponding
to the nodes of the admissible region of one of the two attributables,
let us say ${\cal A}_1$. Then we compute, from each of the $X_i$, a
prediction ${\cal A}_i$ for the epoch $t_2$, each with its covariance
matrix $\Gamma_{{\cal A}_i}$. Thus for each virtual debris $X_i$ we
can compute an attribution penalty $K_4^i$ \cite{orbdet_book} and use
the values as a criterion to select some of the virtual debris to
proceed to the orbit computation.

Thus the procedure is as follows: we select some maximum value
$K_{max}$ for the attribution penalty and if there are some nodes such
that $K_4^i\leq K_{max}$ we proceed to the correlation confirmation.
If this is not the case, we can try with another method, such as the
one described in Sec.~\ref{sec:kepint}.

\subsection{Universal classification of objects}

\begin{figure}[h]
  \includegraphics[width=0.8\textwidth]{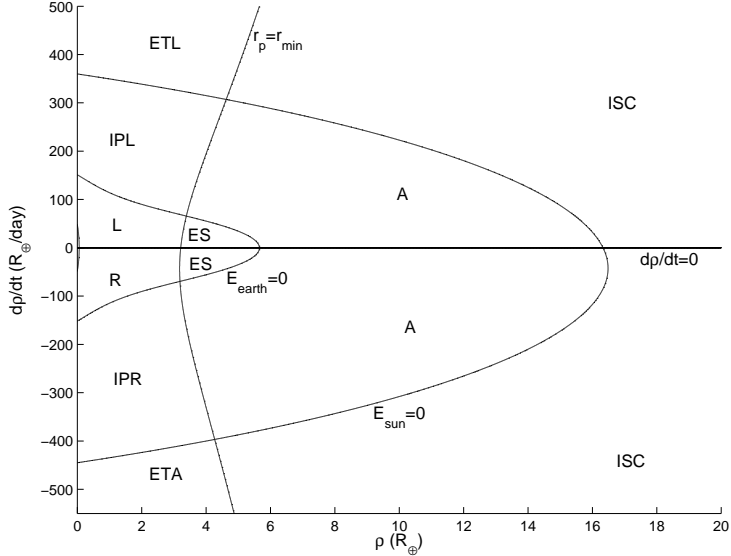}
  \caption{Partitioning of the $(\rho, \dot\rho)$ half plane $\rho>0$
    in regions corresponding to different populations, for an optical
    attributable with proper motion $\eta=10.1980$. The labels mean: L
    Launch, R Reentry, ES Earth Satellite, A Asteroid, ISC
    Interstellar Comet, ETA ET Arriving, ETL ET Leaving, IPR
    Interplanetary Reentry, IPL Interplanetary Launch.  rad/day.}
\label{fig:aregion_univ_optical}
\end{figure}
The method of the admissible region is also useful to provide insight
on the relationship between the different populations, in particular
how they can mix in the observations. For a given optical
attributable, supposedly computed from a short arc of optical
observations, the Fig.~\ref{fig:aregion_univ_optical} shows the region in
the $(\rho, \dot\rho)$ half-plane $\rho>0$ where Earth satellites (ES)
can be, but also where ballistic trajectories (either launches L or
reentries R) can be, and where an asteroid serendipitously found in
the same observations would be. Other more exotic populations, which
are very unlikely, also have their region in the half plane: e.g.,
there are regions for direct departure/arrival to the Earth from
interstellar space, which we have labeled as ET trajectories.

\begin{figure}[t]
  \includegraphics[width=0.8\textwidth]{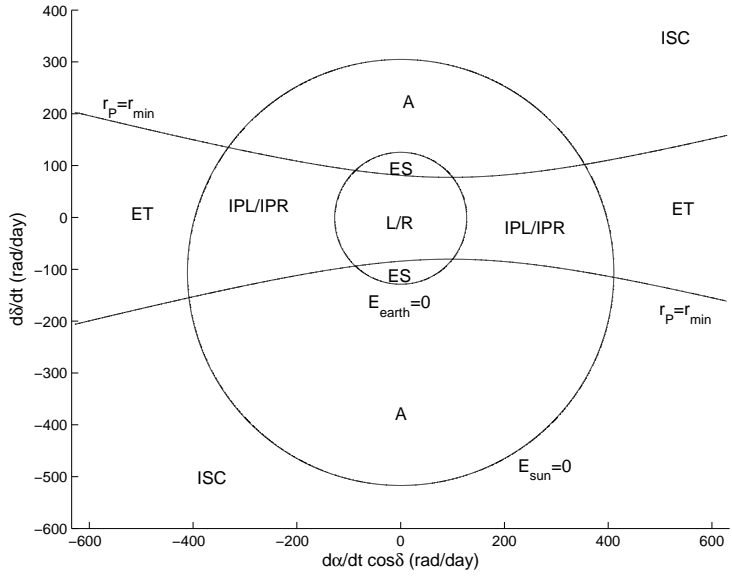}
  \caption{Partitioning of the $(\dot\alpha\cos\delta, \dot\delta)$
    plane in regions corresponding to different populations, for a
    radar attributable with $\rho=1\,R_\oplus$. The labels mean: L/R
    Launch or Reentry, ES Earth Satellite, A Asteroid, ISC
    Interstellar Comet, ET interstellar launch/reentry, IPL/IPR
    Interplanetary Launch or Reentry.}
\label{fig:aregion_univ_radar}
\end{figure}

The same ``universal'' figure can be generated from a given radar
attributable (Fig~\ref{fig:aregion_univ_radar}). In this case the
regions corresponding to different populations partition the plane
$(\dot\alpha\cos\delta, \dot\delta)$.  The curve ${\cal E}_{sun}=0$,
for the heliocentric energy, has been computed with formulas very
similar to the ones for the geocentric energy.

\section{Keplerian integrals method}
\label{sec:kepint}

We shall describe a method proposed for the asteroid case in
\cite{gronchi} and based on the two-body integrals, to produce
preliminary orbits starting from two attributables ${\cal A}_1$,
${\cal A}_2$ of the same object at two epochs $t_1$, $t_2$. We assume
that the orbit between $t_1$ and $t_2$ is well approximated by a
Keplerian 2-body orbit, with constant energy ${\cal E}$ and angular
momentum vector $\mathbf c$:
\begin{equation}\label{keplerian_system}
\begin{cases}
{\cal E}(t_1)-{\cal E}(t_2)=0\\
\mathbf c(t_1)-\mathbf c(t_2)=0
\end{cases}
\, .
\end{equation}

\subsection{Optical case}

Using (\ref{angmom_opt}) and by scalar product between with the first
equation of (\ref{keplerian_system}) and $\mathbf D_1\times\mathbf
D_2$ we obtain the scalar equation of degree 2:
\[
(\mathbf D_1\times\mathbf D_2)\cdot (\mathbf c_1-\mathbf
c_2)=q(\rho_1,\rho_2)=0\, .
\]
Geometrically, this equation defines a conic section in the
$(\rho_1,\rho_2)$ plane.
By the formulae giving $\dot\rho_1$, $\dot\rho_2$ as a function of
$\rho_1$, $\rho_2$ derived from the angular momentum equations:
\begin{eqnarray*}
  \dot\rho_1&=&\frac{(\mathbf E_2\rho_2^2+\mathbf F_2\rho_2+
    \mathbf G_2-\mathbf E_1\rho_1^2-\mathbf F_1\rho_1-\mathbf G_1)
    \times\mathbf D_2}{||\mathbf D_1\times\mathbf D_2||^2}\, ;\\
  \dot\rho_2&=&\frac{\mathbf D_1\times(\mathbf E_1\rho_1^2+\mathbf 
    F_1\rho_1+\mathbf G_1-\mathbf E_2\rho_2^2-\mathbf F_2\rho_2-
    \mathbf G_2)}{||\mathbf D_1\times\mathbf D_2||^2}
\end{eqnarray*}
the energies ${\cal E}_1$, ${\cal E}_2$ can be considered as functions
of $\rho_1$, $\rho_2$ only. Thus we obtain:
\[
\begin{cases}
{\cal E}_1(\rho_1,\rho_2)-{\cal E}_2(\rho_1,\rho_2)=0\\
q(\rho_1,\rho_2)=0
\end{cases}
\, ,
\]
a system of 2 equations in 2 unknowns, already present in \cite{taff}:
they proposed a Newton-Raphson method to solve the system, but this
results into a loss of control on the number of alternate
solutions. In \cite{gronchi} the authors have applied the same
equations to the asteroid problem, and proposed a different approach
to the solution of the system.

The energy equation is algebraic, but not polynomial, because there
are denominators containing square roots. By squaring twice it is
possible to obtain a polynomial equation $p(\rho_1,\rho_2)=0$: the
degree of this equation is 24. Thus the system
\[
\begin{cases}
p(\rho_1,\rho_2)=0\\
q(\rho_1,\rho_2)=0
\end{cases}
\]
has exactly 48 solutions in the complex domain, counting them with
multiplicity. Of course we are interested only in solutions with
$\rho_1$, $\rho_2$ real and positive, moreover the squaring of the
equations introduces spurious solutions.  Nevertheless, we have found
examples with up to 11 non spurious solutions.

We need a global solution of the algebraic system of overall degree
48, providing at once all the possible couples $(\rho_1,\rho_2)$. This
is a classical problem of algebraic geometry, which can be solved with
the resultant method: we can build an auxiliary Sylvester matrix, in
this case $22\times 22$, with coefficients polynomials in $\rho_2$,
and its determinant, the resultant, is a polynomial of degree 48 in
$\rho_2$ only. The values of $\rho_2$ appearing in the solutions of
the polynomial system are the roots of the resultant \cite{cox}.

The computation of the resultant is numerically unstable, because the
coefficients have a wide range of orders of magnitude: we have to use
quadruple precision. Once the resultant is available, there are
methods to solve the univariate polynomial equations, providing at
once all the complex roots with rigorous error bounds
\cite{bini}. Given all the roots which could be real, we solve for the
other variable $\rho_1$, select the positive couples $(\rho_1,\rho_2)$
and remove the spurious ones due to squaring. If the number of
remaining solutions is 0, the attributables cannot be correlated with this
method.

\subsection{Radar case}

The formulae for geocentric energy and angular momentum are given by
(\ref{energy_rad}) and (\ref{angmom_rad}), polynomials of degree 2 and
1 in the unknowns $(\dot\alpha,\dot\delta)$, respectively. The system
(\ref{keplerian_system}) has overall algebraic degree 2: such a system
can be solved by elementary algebra.

The angular momentum equations are
\begin{equation}\label{angmom_eq_rad}
\mathbf A_1\dot\alpha_1 + \mathbf B_1\dot\delta_1+\mathbf C_1=\mathbf
A_2\dot\alpha_2 + \mathbf B_2\dot\delta_2+\mathbf C_2
\end{equation}
that is a system of 3 linear equations in 4 unknowns
$(\dot\alpha_1,\dot\delta_1,\dot\alpha_2,\delta_2)$ and can be solved
for three unknowns as a function of one of the four. For example, by
scalar product between (\ref{angmom_eq_rad}) and $\mathbf
B_1\times\mathbf A_2$ we have
\[
\dot\alpha_1=\frac{\mathbf A_2\cdot(\mathbf B_1\times\mathbf
  B_2)\dot\delta_2-(\mathbf C_1-\mathbf C_2)\cdot(\mathbf
  A_2\times\mathbf B_1)}{\mathbf B_1\cdot(\mathbf A_1\times\mathbf
  A_2)}
\]
and in a similar way we obtain
\begin{eqnarray*}
  \dot\delta_1=\frac{\displaystyle\mathbf B_2\cdot(\mathbf 
    A_1\times\mathbf A_2)\dot\delta_2-(\mathbf C_1-\mathbf 
    C_2)\cdot(\mathbf A_1\times\mathbf A_2)}{\displaystyle
    \mathbf B_1\cdot(\mathbf A_1\times\mathbf A_2)}\ ,\\
  \stackrel{}{
    \dot\alpha_2=\frac{\displaystyle\mathbf A_1\cdot(\mathbf 
      B_1\times\mathbf B_2)\dot\delta_2-(\mathbf C_1-\mathbf 
      C_2)\cdot(\mathbf A_1\times\mathbf B_1)}{\displaystyle
      \mathbf B_1\cdot(\mathbf A_1\times\mathbf A_2)}}\ .
\end{eqnarray*}

When the equations for, say,
$(\dot\alpha_1,\dot\alpha_2,\dot\delta_1)$ as a function of
$\dot\delta_2$ are substituted in the equation for the energies ${\cal
  E}_1(\dot\alpha_1,\dot\delta_1)={\cal
  E}_2(\dot\alpha_2,\dot\delta_2)$ we obtain a quadratic equation in
$\dot\delta_2$, which can be solved by elementary algebra, giving at
most two real solutions. Geometrically, equation (\ref{angmom_eq_rad})
can be described by a straight line in a plane, e.g., in
$(\dot\alpha_2,\dot\delta_2)$, where the energy equation defines a
conic section.

\subsection{Singularities}

There are some cases in which the Keplerian integrals method can not
be applied.

In the optical case we have to avoid the condition $\mathbf
D_1\times\mathbf D_2=(\mathbf q_1\times\hat{\bm\rho}_1)\times(\mathbf
q_2\times\hat{\bm\rho}_2)=0$. This can happen when:
\begin{itemize}
\item $\mathbf q_1$ is parallel to $\hat{\bm\rho}_1$, i.e. the
  observation at time $t_1$ is done at the observer zenith;
\item $\mathbf q_2$ is parallel to $\hat{\bm\rho}_2$, i.e., the
  observation at time $t_2$ is done at the observer zenith;
\item $\mathbf q_1$, $\mathbf q_2$, $\hat{\bm\rho}_1$ and
  $\hat{\bm\rho}_2$ are coplanar. This case arises
  whenever a geostationary object is observed from the same station at
  the same hour of distinct nights.
\end{itemize}
As it is normal, the mathematical singularity is surrounded by a
neighborhood in which the method is possible for zero error (both
zero observational error and zero rounding off in the computation),
but is not applicable in practice. E.g., for nearly geosynchronous
orbits, even if they are not geostationary, and for hours of
observations in different nights different by few minutes, this method
fails.

In the radar case the procedure fails only if the four vectors
$\mathbf A_1$, $\mathbf A_2$, $\mathbf B_1$ and $\mathbf B_2$ do not
generate a linear space of dimension 3, i.e., when:
\[
\begin{cases}
\mathbf A_i\cdot(\mathbf B_1\times\mathbf B_2)=0\\
\mathbf B_i\cdot(\mathbf A_1\times\mathbf A_2)=0
\end{cases}
i=1,2.
\]
For $i=1$ we obtain
\[
\begin{cases}
  \rho_1^2\rho_2[\hat{\bm\rho}_{\delta 2}\cdot(\mathbf
  r_1\times\mathbf r_2)][\mathbf r_1\cdot(\hat{\bm\rho}_{\alpha
    1}\times\hat{\bm\rho}_{\delta 1})]=0\\
  \rho_1^2\rho_2[\hat{\bm\rho}_{\alpha 2}\cdot(\mathbf
  r_1\times\mathbf r_2)][\mathbf r_1\cdot(\hat{\bm\rho}_{\delta
    1}\times\hat{\bm\rho}_{\alpha 1})]=0
\end{cases}
\]
and for $i=2$ the formulae are analogous.
Thus there is singularity when:
\begin{itemize}
\item $\mathbf r_1$ is parallel to $\mathbf r_2$;
\item $\mathbf r_i\cdot(\hat{\bm\rho}_{\delta
    i}\times\hat{\bm\rho}_{\alpha i})=\cos\delta_i(\mathbf
  q_i\cdot\hat{\bm\rho}_i+\rho_i)=0$, but this can never happen, apart
  from coordinate singularities, because $\mathbf
  q_i\cdot\hat{\bm\rho}_i\geq 0$;
\item $\mathbf r_1\cdot(\hat{\bm\rho}_{\delta
    1}\times\hat{\bm\rho}_{\alpha i})=0$ and $\mathbf
  r_1\cdot(\hat{\bm\rho}_{\delta 1}\times\hat{\bm\rho}_{\delta i})=0$,
  i.e., $\hat{\bm\rho}_{\alpha i}$ and $\hat{\bm\rho}_{\delta i}$ for
  $i=1,2$ belong to the orbital plane.
\end{itemize}

\subsection{Preliminary orbits}

Once a solution of (\ref{keplerian_system}) is computed the values of
attributable elements can be obtained for the epochs $\bar t_1$ and $\bar t_2$,
and they can be converted into the usual Keplerian elements:
\[
(a_j,e_j,I_j,\Omega_j,\omega_j,\ell_j)\ ,\ j=1,2\ ,
\]
where $\ell_j$ are the mean anomalies. The first four Keplerian
elements $(a_j,e_j,I_j,\Omega_j)$ are functions of the 2-body energy
and angular momentum vectors ${\cal E}_j$, $\mathbf c_j$, and are the
same for $j=1,2$. Thus the result can be assembled in the
8-dimensional vector:
\begin{equation}\label{vector8}
  H=(V,\Phi_1,\Phi_2) \ \ ,\ \
  V=(a,e,I,\Omega)\ , \ \Phi_1=(\omega_1,\ell_1) \ ,\
  \Phi_2=(\omega_2,\ell_2) \ .
\end{equation}
There are compatibility conditions between $\Phi_1$ and $\Phi_2$ to be
satisfied if the two attributables belong to the same object:
\begin{equation}\label{compatibility_conditions}
\omega_1=\omega_2\ ,\ \ell_1=\ell_2+n(\bar t_1-\bar t_2)\ ,
\end{equation}
where $n=n(a)$ is the mean motion.  We cannot demand the exact
equality in the formulae above, because of various error sources,
including the uncertainty of the attributable, and the changes on the
Keplerian integrals due to the perturbations with respect to the
2-body model. Thus we need a metric to measure in an objective way the
residuals in the compatibility conditions.

\subsection{Covariance propagation}

The two attributables ${\cal A}_1,{\cal A}_2$ used to compute the
coefficients of equations (\ref{keplerian_system}) have been computed
from the observations by using a least squares fit to the individual
observations, thus $4\times 4$ covariance matrices $\Gamma_{{\cal
    A}_1}$ and $\Gamma_{{\cal A}_2}$ are available; they can be used
to form the block diagonal $8\times 8$ covariance matrix for both
attributables $\Gamma_{\cal A}$. The Keplerian integral method allows
to compute explicitly the vector $H$ of (\ref{vector8}) and, by means
of the implicit function theorem, its partial derivatives, thus it is
possible by the standard covariance propagation formula
\cite{orbdet_book}[Sec. 5.5] to compute also $\Gamma_H$, the
covariance of $H$. With another transformation we can compute the
average elements $\Phi_0=(\Phi_1+\Phi_2)/2$ (as the best value for the
angular elements at time $\bar t_0=(\bar t_1+\bar t_2)/2$) and the
discrepancy $\Delta\Phi$ in the compatibility conditions
(\ref{compatibility_conditions}), and to propagate the covariance also
to this 8-dimensional vector:
\[
\Gamma_{\cal A} \Longrightarrow \Gamma_H \Longrightarrow
\Gamma_{V,\Phi_0,\Delta\Phi}\ .
\]
The above argument is a generalization of the one in \cite{gronchi},
where explicit computations are given for the optical attributables
case.

In the $8\times 8$ covariance matrix $\Gamma_{V,\Phi_0,\Delta\Phi}$,
the lower right $2\times 2$ block is the marginal covariance matrix of
$\Delta\Phi$, from which we can compute the normal matrix and the
$\chi^2$:
\[
C_{\Delta\Phi}=\Gamma^{-1}_{\Delta\Phi}\ \ , \ \ \chi^2_{\Delta\Phi}=
\Delta\Phi\cdot C_{\Delta\Phi}\,\Delta\Phi\ ,
\]
which can be used as control, that is the discrepancy in the
compatibility conditions is consistent with the observation error and
the correlation between the two attributables is considered possible
only if $\chi^2_{\Delta\Phi}\leq \chi^2_{max}$.

The upper left $6\times 6$ block is the covariance matrix of the
preliminary orbit, that is of the orbital elements set $(V, \Phi_0)$
(at epoch $\bar t_0$). Although this preliminary orbit is just a
2-body solution, it has an uncertainty estimate, arising from the
(supposedly known) statistical properties of the observational
errors. This estimate neglects the influence of perturbations, such as
the spherical harmonics of the Earth gravity field, the lunisolar
differential attraction and the non-gravitational perturbations;
nevertheless, if the time span $\bar t_2-\bar t_1$ is short, the
covariance obtained above can be a useful approximation.

\subsection{Precession model}

We can generalize the method, including the effect due to the
non-spherical shape of the Earth. The averaged equation for Delaunay's
variables $\ell$, $g=\omega$, $z=\Omega$, $L=\sqrt{\mu a}$,
$G=L\sqrt{1-e^2}$ and $Z=G\cos I$ are \cite{roy}[Sec. 10.4]:
\begin{equation}\label{delaunay_eq}
\begin{cases}
  \displaystyle\bar{\dot\ell}=n-\frac{3}{4}n
  \left(\frac{R_\oplus}{a}\right)^2\frac{J_2(1-3\cos^2I)}{(1-e^2)^{3/2}}\\
  \stackrel{}{\displaystyle\bar{\dot
      g}=\frac{3}{4}n\left(\frac{R_\oplus}
      {a}\right)^2\frac{J_2(4-5\sin^2I)}{(1-e^2)^2}}\\
  \stackrel{}{\displaystyle\bar{\dot
      z}=-\frac{3}{2}n\left(\frac{R_\oplus}
      {a}\right)^2\frac{J_2\cos I}{(1-e^2)^2}}\\
  \stackrel{}{\bar{\dot L}=\bar{\dot G}=\bar{\dot Z}=0}
\end{cases}\ ,
\end{equation}
where $J_2$ is the coefficient of the second zonal spherical harmonic
of the Earth gravity field. To apply in this case the Keplerian
integrals method, we can not use the equations assuming conservation
of the angular momentum. From (\ref{delaunay_eq}) we can replace
(\ref{keplerian_system}) with:
\begin{equation}\label{j2_system}
\begin{cases}
  {\cal E}_1={\cal E}_2\\
  \mathbf c_1\cdot\hat{\mathbf z}=\mathbf c_2\cdot\hat{\mathbf z}\\
  ||\mathbf c_1||^2=||\mathbf c_2||^2\\
  \cos(z_2)=\cos(z_1)\cos(\bar{\dot z}(\bar t_2-\bar
  t_1))-\sin(z_1)\sin(\bar{\dot z}(\bar t_2-\bar t_1))
\end{cases}
\, .
\end{equation}
In the optical case the first equation is algebraic and by squaring
twice is possible to obtain a polynomial equation; in the radar case
this relation is already polynomial. The second and the third
equations are always polynomial, while the last equation needs to be
linearized in the parameter $\bar{\dot z}(\bar t_2-\bar t_1)$:
\begin{equation}\label{eq_cos_z}
\cos z_2=\cos z_1-\bar{\dot z}(\bar t_2-\bar t_1)\sin z_1\, .
\end{equation}
The following relationships hold:
\begin{align*}
  \cos z_i&=\frac{\hat{\mathbf z}\times\mathbf
    c_i}{||\hat{\mathbf z}\times\mathbf c_i||}\cdot\hat{\mathbf x}\,
  ,&\ \ \sin z_i&=\frac{\hat{\mathbf z}\times\mathbf 
    c_i}{||\hat{\mathbf z}\times\mathbf c_i||}\cdot\hat{\mathbf y}\, ,
  &\ \  a(1-e^2)&=\frac{||\mathbf c_1||^2}{\mu}\, ,\\
  \cos I&=\frac{\mathbf c_1\cdot\hat{\mathbf 
      z}}{||\mathbf c_1||}\, ,&\ \ n&=\sqrt{-\frac{8{\cal 
	E}_1^3}{\mu^2}}\, ,&\ \ \bar{\dot z}(t_2-t_1)
  &=\frac{\xi\sqrt{-8{\cal E}_1^3}\,(\mathbf c_1\cdot \hat{\mathbf 
      z})}{||\mathbf c_1||^5}
\end{align*}
where $\xi=-3\mu J_2 R_\oplus^2(\bar t_2-\bar t_1)/2$. Substituting in
(\ref{eq_cos_z}) we obtain
\[
\frac{\hat{\mathbf z}\times\mathbf c_2}{||\hat{\mathbf z}\times\mathbf
  c_2||}\cdot\hat{\mathbf x}=\frac{\hat{\mathbf z}\times\mathbf
  c_1}{||\hat{\mathbf z}\times\mathbf c_1||}\cdot\hat{\mathbf
  x}-\frac{\xi\sqrt{-8{\cal E}_1^3}\,(\mathbf c_1\cdot \hat{\mathbf
    z})}{||\mathbf c_1||^5}\frac{\hat{\mathbf z}\times\mathbf
  c_1}{||\hat{\mathbf z}\times\mathbf c_1||}\cdot\hat{\mathbf y}\, .
\]
Since $||\hat{\mathbf z}\times\mathbf c||=||\mathbf c||\sin I$ is
constant we have:
\[
||\mathbf c_1||^5[\hat{\mathbf z}\times(\mathbf c_2-\mathbf
c_1)]\cdot\hat{\mathbf x}=-\xi\sqrt{-8{\cal E}_1^3}\,(\mathbf c_1\cdot
\hat{\mathbf z})(\hat{\mathbf z}\times\mathbf c_1)\cdot\hat{\mathbf 
  y}\ ,
\]
that is an algebraic equation. Furthermore, by squaring twice in the
optical case and only once in the radar case it is possible to obtain
a polynomial equation.

Finally the new compatibility conditions, in place of
(\ref{compatibility_conditions}) need to take into account the
precession of the perigee and the secular perturbation in mean
anomaly:
\[
g_1=g_2+\bar{\dot g}(\bar t_1-\bar t_2)\, ,\ \
\ell_1=\ell_2+\bar{\dot\ell}(\bar t_1-\bar t_2)\, .
\]
The overall degree of system (\ref{j2_system}) is summarized in Table
\ref{tab:degree}. We conclude that this method is unpractical
for optical attributables, could be used for radar attributables,
with computational difficulties comparable with the optical case
without precession.
\begin{table}[h]
  \caption{Degrees of the equations in system (\ref{j2_system})}
\label{tab:degree}
\begin{tabular}{lll}
  \hline\noalign{\smallskip}
  & Optical case & Radar case  \\
  \noalign{\smallskip}\hline\noalign{\smallskip}
  ${\cal E}_1={\cal E}_2$ & 16  & 2 \\
  $\mathbf c_1\cdot\hat{\mathbf z}=\mathbf c_2\cdot\hat{\mathbf 
    z}$ & 2 & 1 \\
  $||\mathbf c_1||^2=||\mathbf c_2||^2$ & 4 & 2 \\
  $\cos z_2=\cos z_1-\bar{\dot z}(\bar t_2-\bar 
  t_1)\sin z_1$ & 54 & 12 \\
  Total & 6912 & 48 \\
  \noalign{\smallskip}\hline
\end{tabular}
\end{table}

To solve the problem (even in the optical case) we begin by
considering the parametric problem $\bar{\dot
  z}=K$, where $K$ is constant. Thus we replace
(\ref{keplerian_system}) with:
\[
\begin{cases}
  {\cal E}_1-{\cal E}_2=0\\
  R\,\mathbf c_1-R^T\,\mathbf c_2=0
\end{cases}
\]
where $R$ is the rotation by $\Delta\Omega/2=K(\bar t_2-\bar t_1)/2$
around $\hat{\mathbf z}$.  This means that for a fixed value of $K$
the problem has the same algebraic structure of the unperturbed
one. The only thing needed is to substitute $\mathbf D_1$, $\mathbf
E_1$, $\mathbf F_1$ and $\mathbf G_1$ with $R\mathbf D_1$, $R\mathbf
E_1$, $R\mathbf F_1$ and $R\mathbf G_1$ in the optical case and
$\mathbf A_1$, $\mathbf B_1$ and $\mathbf C_1$ with $R\mathbf A_1$,
$R\mathbf B_1$ and $R\mathbf C_1$ in the radar case; similarly the
vectors with index 2 are multiplied by $R^T$.

The compatibility conditions contain the precession of
the perigee and the secular perturbation in mean anomaly,
related to the one of the node by linear equations
\[
g_1=g_2+K C_g\,(\bar t_1-\bar t_2)\, ,\ \
\ell_1=\ell_2+(n+K C_\ell)(\bar t_1-\bar t_2)\, ,
\]
where the coefficients $C_g, C_\ell$ can be easily deduced from
(\ref{delaunay_eq}). Thus we can compute the $\chi^2_{\Delta\Phi}(K)$
and set up a simple procedure to minimize this by changing $K$, then
the control on the acceptability of the preliminary orbit is
\[
\min_K \chi^2_{\Delta\Phi}(K)\leq \chi^2_{max}\ .
\]

\section{Correlation confirmation}
\label{sec:correl_confirm}

The multiple orbits obtained from the solutions of the algebraic
problem are just preliminary orbits, solution of a 2-body
approximation (as in the classical methods of Laplace and Gauss), or
possibly of a $J_2$-only problem. They have to be replaced by least
squares orbits, with a dynamical model including all the relevant
perturbations.

Even after confirmation by least squares fit, it might still be the
case that some linkages with just two attributables can be
\emph{false}, that is the two attributables might belong to different
objects. This is confirmed by the tests with real data reported in
\cite{tommei_esoc} for the Virtual debris method and in
\cite{milani_esoc} for the Keplerian integrals method. \cite{gronchi}
have found the same phenomenon in a simulation of the application of
the same algorithm to the asteroid case. Thus every linkage of two
attributables needs to be confirmed by correlating a third
attributable.

The process of looking for a third attributable which can also be
correlated to the other two is called attribution
\cite{milani_recovery,milani_attrib}. From the available
2-attributable orbit with covariance we predict the attributable
${\cal A}_P$ at the time $t_3$ of the third attributable, and compare
with ${\cal A}_3$ computed from the third set of
observations. Both ${\cal A}_P$ and ${\cal A}_3$ come with a
covariance matrix, we can compute the $\chi^2$ of the difference and
use it as a test. For the attributions passing this test we proceed to
the differential corrections.
The procedure is recursive, that is we can use the 3-attributable
orbit to search for attribution of a fourth attributable, and so
on. This generates a very large number of many-attributable orbits,
but there are many duplications, corresponding to adding them in a
different order.

By correlation management we mean a procedure to remove duplicates
(e.g., $A=B=C$ and $A=C=B$) and inferior correlations (e.g., $A=B=C$
is superior to both $A=B$ and to $C=D$, thus both are removed). The
output catalog after this process is called normalized. In the
process, we may try to merge two correlations with some attributables
in common, by computing a common orbit \cite{milani_id}.

\section{Conclusions}

We have described two algorithms to solve the linkage problem, that is
to compute an orbit for an Earth-orbiting object observed in two well
separated arcs. 
The first method exploits the geometric structure of the admissible
region of negative geocentric energy orbits, which is sampled to
generate virtual orbits. The latter are propagated in time to find
other observations which could belong to the same object.
The second method exploits the integrals of the 2-body problem, which
are constant even over a significant time span and thus should apply
to both observed arcs of the same object.

This top level description is enough to understand that the Virtual
debris algorithm should be applied to short time intervals between
observed arcs, less than one orbital period or at most a few orbital
periods. The Keplerian integrals method can be used for longer time
spans, spanning several orbital periods; it is near to a singularity
for very short time spans and in some other near-resonance conditions,
such as observations of a geosynchronous orbits at the same hour in
different nights. 
We conclude that each method should be used in the cases in which it
is most suitable. Both algorithms have been tested for the optical
case with real data from the ESA Optical Ground Station
\cite{tommei_esoc,milani_esoc} with good results. The analogous
algorithms have been tested for asteroids in simulations of next
generation surveys \cite{milani_id,gronchi}. Future work should
include the tests of the radar case and the solution of other related
problem, like \emph{orbit identification} between two objects for
which an orbit is already available.

\begin{acknowledgements}
Part of this work was performed in the framework of  ESOC Contract No.
21280/07/D/CS, ``Orbit Determination of Space Objects
Based on Sparse Optical Data''.
\end{acknowledgements}

\end{document}